\documentclass[letterpaper]{iopart}
\usepackage{graphicx}
\usepackage[dvipdfm]{color}
\usepackage{ulem}
\usepackage{amssymb}
\usepackage{latexsym}

\begin{document}
 
\title{Search method for coincident events from LIGO and IceCube
detectors}
\author{Yoichi Aso$^1$, Zsuzsa M\'{a}rka$^1$, Chad Finley$^2$, John
Dwyer$^1$, Kei Kotake$^3$, Szabolcs M\'{a}rka$^1$} 

\address{$^1$Department of Physics, Columbia University, New York, NY,
10027, USA} 
\address{$^2$Department of Physics, University of Wisconsin, Madison,WI, 53706, USA}
\address{$^3$Division of Theoretical Astronomy, National Astronomical
Observatory of Japan, Mitaka, Tokyo, Japan}
 \ead{aso@astro.columbia.edu}

\begin{abstract}
 We present a coincidence search method for astronomical events using
 gravitational wave detectors in conjunction with other astronomical
 observations. We illustrate our method for the specific case of
 the LIGO gravitational wave detector and the IceCube
 neutrino detector.    LIGO trigger-events and
 IceCube events which  occur  within a  given time window
 are selected as time-coincident events.  Then the spatial overlap of
 the reconstructed event directions is evaluated
  using an unbinned maximum likelihood
 method. Our method was tested  with Monte
 Carlo simulations  based on
 realistic LIGO and IceCube event
 distributions. We estimated a typical false alarm rate
  for the analysis to be 1 event per 435
 years.  This
 is significantly smaller than the false alarm rates of the individual detectors.

\end{abstract}
\pacs{95.55.Ym, 95.55.Vj, 04.80.-y}
\submitto{\CQG}
\section{Introduction}
In this paper, we present an analysis method to look for astrophysical
sources which produce both gravitational wave and high energy neutrino
bursts using data from LIGO and IceCube.  One example of
 a possible source
is a gamma-ray burst (GRB).  Thanks to the {\it Swift}
satellite\,\cite{2004ApJ...611.1005G}, there is accumulating
observational evidence suggesting the association of long GRBs with the
death of massive stars and supernova-like events (e.g. SN2006aj and
GRB060218\,\cite{2006Natur.442.1008C}, see
also\,\cite{2005tmgm.meet..860L}).  The collapsar
model\,\cite{1999A&AS..138..499W} is widely accepted for explaining long
GRBs and stellar collapse.  During the gravitational collapse of rapidly
rotating stars, gravitational waves are emitted (see
\cite{2006RPPh...69..971K} for a review).  First,
fireballs heated by neutrinos from the accretion disk are
 thought to produce the prompt
gamma-ray emissions \,\cite{2005RvMP...76.1143P}. Subsequently
in the prompt and afterglow phases, high energy neutrinos
($\sim 10^5-10^{10}$\,GeV) are expected to be produced
by accelerated protons in relativistic shocks (see
\cite{1997PhRvL..78.2292W, 1998PhRvL..80.3690V} for reviews).  High
energy neutrinos could be emitted also from short-duration GRBs, which
are thought to be the outcome of neutron star
mergers\,\cite{2007NJPh....9...17L}.  There is currently limited
knowledge both observationally and theoretically about the details of
the astrophysical process connecting  the gravitational collapse/merger of
compact objects and  black hole
formation  with the formation of
fireballs. Coincident observations of gravitational waves and neutrinos
from those events could therefore make an important contribution to the
understanding of such phenomena.

Apart from GRBs, there may be other (unknown) classes of sources which
produce bright bursts both in gravitational waves and neutrinos. Since
our proposed method is not specific to any source type, our search will
be able to set an upper limit for the population of any sources that
produce nearly simultaneous bursts of gravitational waves and high energy neutrinos
within the detection range of LIGO and IceCube.  We may also discover a
previously unknown astrophysical phenomenon,  if correlated events are found
at a high confidence level.

There are several interferometric gravitational wave
(GW)\,\cite{2001astro.ph.10349H} detectors around the world, such as
LIGO\,\cite{2006CQGra..23S..51S}, TAMA\,\cite{2004CQG..21..403T},
GEO\,\cite{2006CQGra..23S..71L} and VIRGO\,\cite{2006CQGra..23S..63A},
currently in operation.  These detectors
monitor the relative displacement of mirrors (test masses) in response
to distortions induced by gravitational waves. There are
also several high-energy neutrino detectors operating, including
AMANDA\,\cite{ahrens_search_2004}, IceCube\,\cite{2004APh....20..507A},
and ANTARES\,\cite{1999astro.ph..7432A}, which look for the Cherenkov light of charged particles
emitted by neutrino interactions in water or ice. We illustrate the
coincidence search method for the specific case of LIGO and IceCube.

LIGO is a network of interferometric gravitational wave detectors
consisting of three interferometers\footnote{From now
on, we treat the network of the three LIGO interferometers as one
detector and use the word ``detector'' to refer to them as a whole. To refer
to individual LIGO interferometers, we always use the word
``interferometer'' to avoid confusion.} in the USA\,\cite{1992Sci...256..325A}. Two
interferometers (4\,km and 2\,km long ones) are co-located in Hanford,
WA and another 4\,km interferometer is located in Livingston, LA. They
have now achieved the design sensitivity\,\cite{2007AIPC..928...11F}.

Since the interaction of gravitational waves with matter is extremely
weak, expected signals even from very strong gravitational wave sources
are very small.  In order to declare a detection, we have to find a
small signal in an overwhelming noise
background  with high
confidence. Generally, the output from the detector contains glitches
which are not associated with gravitational waves but rather caused by
various local disturbances such as laser noises, seismic excitations,
etc. In order to search for GW bursts, which are gravitational waves of
short duration, it is therefore important to distinguish gravitational
wave signals from noise glitches without prior knowledge of signal
waveforms.

One way to pick  out gravitational wave
signals of unknown waveform from the noise background is to find
coincidences between independent detectors. We can reject a large
 fraction of background events by
comparing the arrival time and other properties (frequency, duration,
etc) of the signals detected by independent gravitational wave
detectors\,\cite{2006CQGra..23S..29A, 2004PhRvD..69j2001A,
2005PhRvD..72l2004A}.  Additionally, event lists from other astronomical
observations, such as Gamma-Ray Bursts (GRB), optical supernovae,
neutrinos, etc., can be used to find events that may be associated with
GW bursts with an increased likelihood\,\cite{2006AIPC..836..605M,
2007PRD..76S..062003T, Marka:2005nz, Mohanty:2004tk,
2004APh....21..201A}.  Moreover, strict coincidence requirements allow
us to investigate candidate events at lower Signal-to-Noise Ratios
(SNRs) while maintaining a low False Alarm Rate (FAR).  Here, we propose
a method for coincidence analysis of gravitational wave data with other
detectors and illustrate it for the case of the LIGO gravitational wave
detector and the IceCube neutrino detector.

IceCube is a cubic kilometer-scale
neutrino detector under construction at the geographic South Pole. Its
primary mission is the search for high energy extraterrestrial
neutrinos. When completed, IceCube will consist of an array of 4800 digital
optical modules, attached to 80 strings submerged within the Antarctic
ice.  Currently the detector is taking data with more than 90\,\%
livetime, except during a few months each year for construction and
commissioning of new strings.  IceCube is optimized to look "down,"
using the Earth as a screen to block all particles except neutrinos;
thus its field of view is the northern hemisphere.  Neutrino arrival
directions are resolved with a median error between 1$^\circ$ and
2$^\circ$\,\cite{2007ICRC...IPC}. The threshold neutrino
energy for the IceCube detector is ~ 100\,GeV.  The full energy range of
observed events depends primarily on the competition between the
unknown, but presumably falling, source flux, versus the rising neutrino
cross-section. A flux with an $E^{-2}$ differential energy spectrum, for
example, results in an energy distribution of neutrino events that
peaks in the range of $10^4$ to $10^5$\,GeV.

In our search method, the data streams from the LIGO interferometers are
processed by a trigger generation pipeline, which generates a
 list of gravitational wave triggers for
each interferometer. Then we compare the trigger lists from LIGO
interferometers to generate a coincident LIGO event list, which contains
the arrival time and the source direction of each  event. The
LIGO event list is compared with an event list from the IceCube detector
which also contains the timing and source direction information of the
events. From the event lists we choose pairs of LIGO-IceCube events
which lie within a certain time interval as time-coincident events. Then
the spatial overlap between the LIGO and IceCube events is statistically
evaluated to obtain the significance of the coincident event.

Because of the very different nature and geographical location of the
two detectors, it is extremely unlikely that  the coincident triggers are due to the same
source of noise. Therefore, the remaining possibility for time
coincident trigger generation in both detectors, other than real
astronomical events, is accidental coincidence.  Furthermore, the chance
for two time-coincident noise triggers to generate overlapping
reconstructed directions on the sky is also  small.  By the
combination of  timing and directional coincidence discrimination, we
can expect that most background events will be rejected and the FAR will
be significantly reduced.

\section{Coincidence analysis}
\begin{figure}[tbp]
 \begin{center}
  \includegraphics[width=12cm]{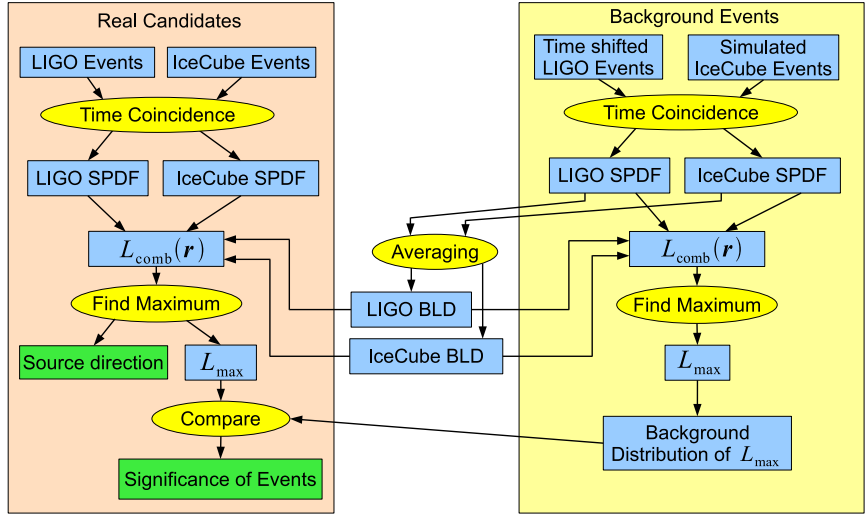} \caption{Outline
of the analysis pipeline. SPDF: Spatial Probability Distribution
Function. BLD: Background Likelihood Distribution. $L_\mathrm{comb}(\bi{r})$:
Combined Likelihood Distribution. $L_\mathrm{max}$: The maximum value of
$L_\mathrm{comb}$.} \label{FlowChart}
 \end{center}
\end{figure}

The outline of the proposed analysis method is shown in
figure\,\ref{FlowChart}. The inputs to the analysis pipeline are LIGO
and IceCube event  lists and a large
number of simulated background events. The outputs of the pipeline are
the most plausible source direction and the statistical significance of
 any time-coincident event against
the background noise events.

\subsection{Event lists}
Data streams from LIGO interferometers are processed by a trigger
generation pipeline (e.g.\,\cite{Beauville05-2, Chatterji04}) to generate
a list of events  for each LIGO interferometer. We then compare
the arrival times of the events from the LIGO interferometers and select
events which appear in all the detectors with less than 10\,ms time
difference. 10\,ms corresponds to the gravitational wave's travel time
between the two LIGO sites, i.e.  the maximum time delay allowed for a
gravitational wave signal.  If the trigger generation pipeline provides
more information on the  events, such as dominant frequency,
duration, etc., we also compare those parameters and reject events with
large discrepancies.

This intra-LIGO coincidence can be applied between all three LIGO
interferometers or any combination of two interferometers. From now on
in this paper, we focus on the two-interferometer case using the
Hanford 4\,km (H1) and the Livingston 4\,km (L1) interferometers, because
the third interferometer (Hanford 2\,km) is two times less sensitive than
the others.

For later statistical treatments, a large number of background events
are created, also from the LIGO data, in almost the same way. The only
difference is that we introduce an artificial time shift between the
trigger times from different interferometers to ensure that the resultant
background event list does not contain real gravitational wave events.

An IceCube event list is determined by the combination of event
reconstruction algorithms and quality cuts used to reject the dominant
background of down-going cosmic ray muons. The remaining up-going events
are expected to be predominantly atmospheric neutrinos, produced by
cosmic rays on the far side of the Earth. The individual event
information needed for this analysis is the time, the arrival direction,
and its associated angular uncertainty. For background IceCube events,
Monte Carlo simulations which imitate the distributions and average
properties of IceCube events have been used.
\subsection{Time coincidence}
Once event lists from LIGO and IceCube are prepared, they are compared
for inter-detector time coincidence. We look for pairs of LIGO and
IceCube events which appear within a certain time window and register
them as time-coincident  events for
further analysis.

A smaller time window can reject background events more efficiently.
However, the size of the time window must be sufficiently large to allow
intrinsic time delay between the two emission processes at the source.
Since we do not assume any specific source model in this analysis, we
propose to use several time windows e.g. 0.1, 1, 10\,sec and also 1\,day
in the case of long GRB search. The time window should be
larger than the travel time of light between the IceCube and LIGO sites,
i.e. 40\,msec.

\subsection{Spatial coincidence}
\begin{figure}[htbp]
\begin{center}
\begin{minipage}[c]{5cm}
\begin{center}
\includegraphics[height=5cm]{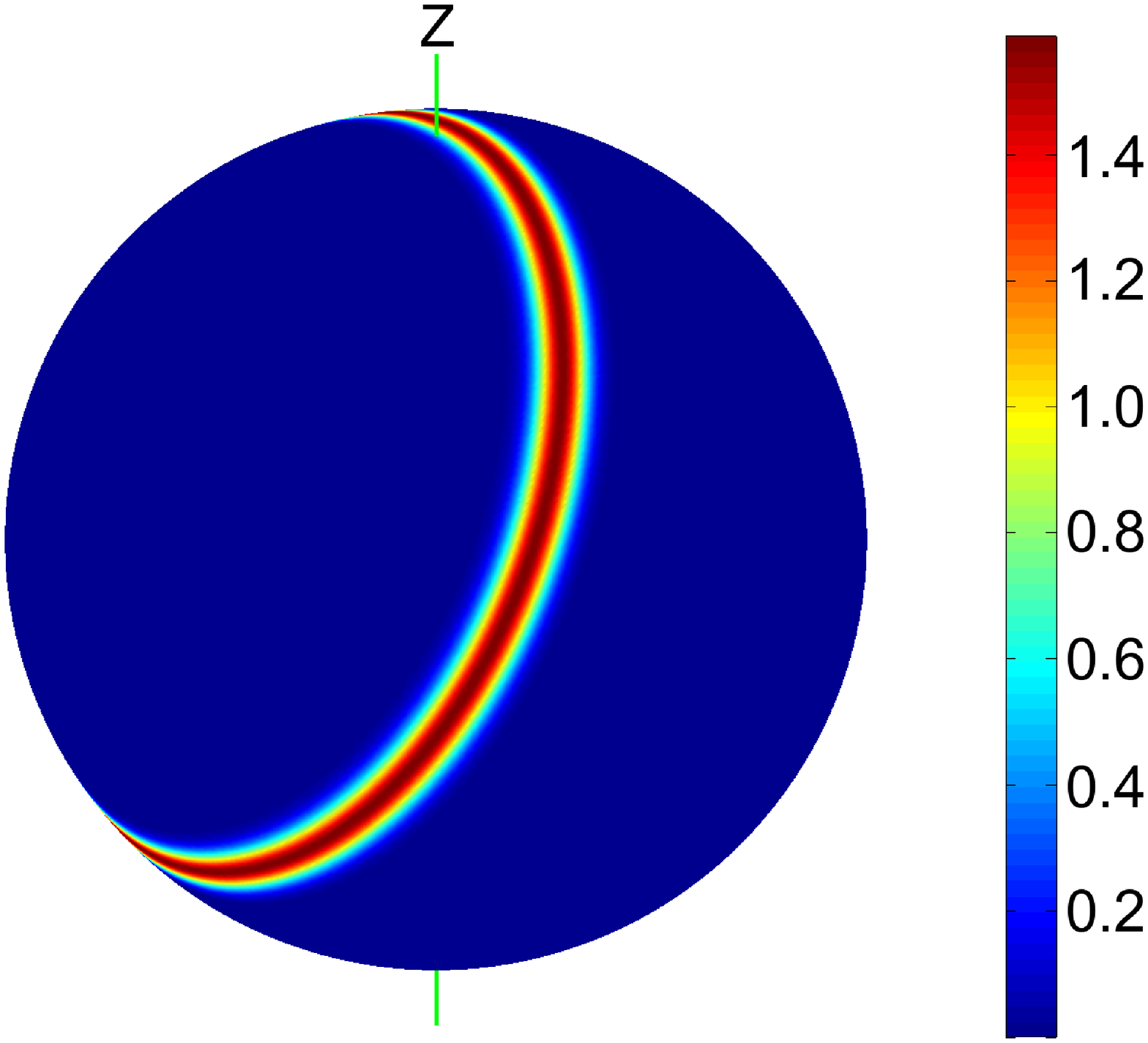} \\
(a)
\end{center}
\end{minipage}
\hspace{1.5cm}
\begin{minipage}[c]{5cm}
\begin{center}
\includegraphics[height=5cm]{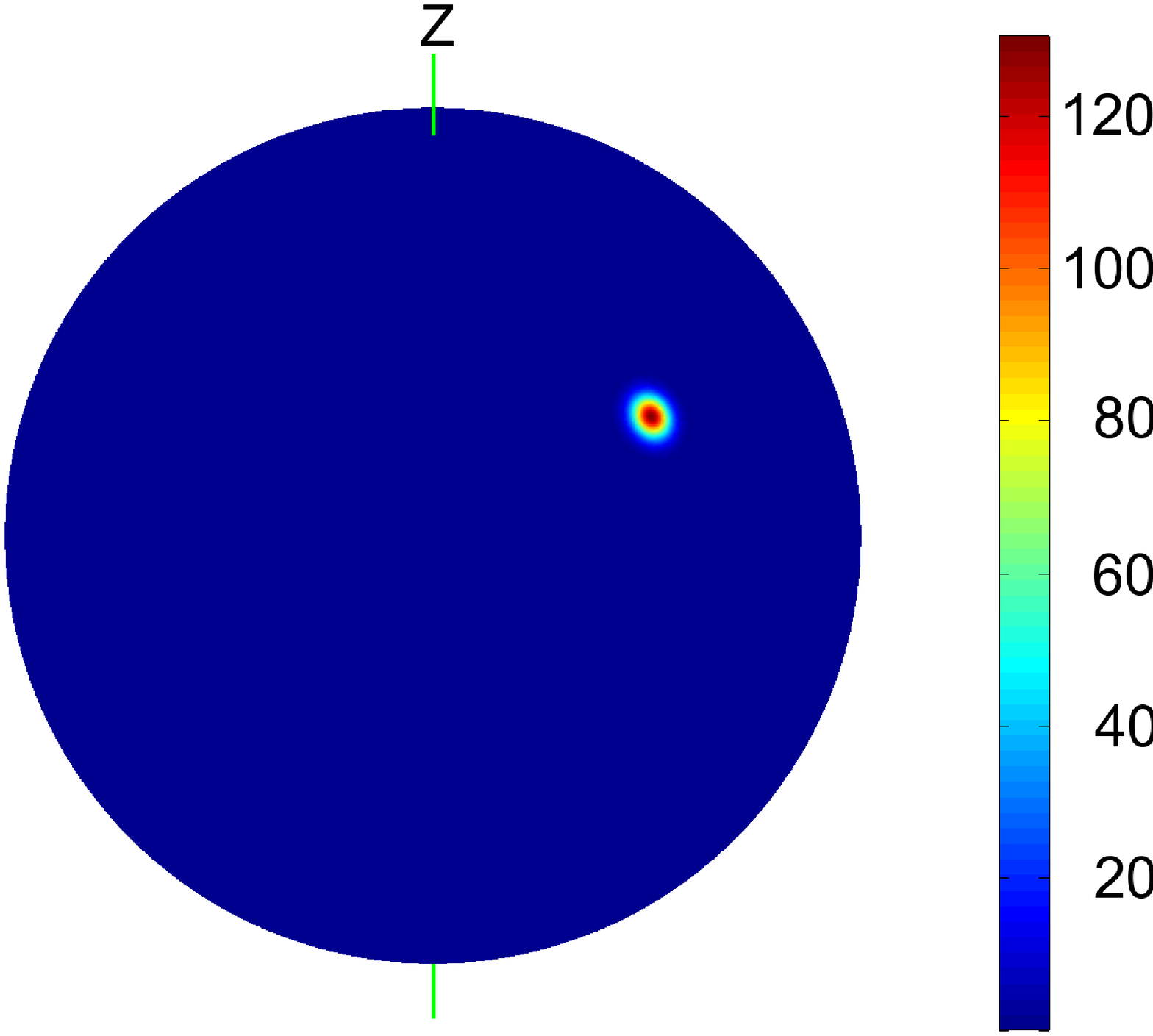} \\
(b)
\end{center}
\end{minipage}\\
\caption{Examples of spatial probability distribution functions
 (SPDFs). (a) SPDF of a LIGO event with $\tau=4$\,msec and
 $\delta\tau=440\,\mu$sec. (b) SPDF of an IceCube event with
 $\sigma_\nu=2^\circ$. The plots are shown in Earth based coordinates
 with the z-axis pointing along the north pole. Both SPDFs are
 normalized to 1 for integration over the sphere.} \label{figure: SPDF}
\end{center}
\end{figure}

The LIGO-IceCube combined events which survive the time-coincidence
discrimination are further processed in order to examine spatial coincidence by
an unbinned maximum likelihood method. 

First, we calculate the Spatial Probability Distribution Function (SPDF)
of each event from LIGO and IceCube. Taking a sky location $\bi{r}$
as an input, this function returns the probability of the actual source
location being $\bi{r}$.

The source location of each LIGO event is reconstructed by measuring the
arrival time difference $\tau$ of the signal between the two
sites. Using the measured arrival time difference $\tau_\mathrm{M}$, we
can constrain the possible source locations to a ring on the sky defined
by a polar angle $\theta_\mathrm{ev}=\cos^{-1}\left(c
\tau_\mathrm{M}/D\right)$ measured from the axis connecting the two LIGO
sites (LIGO axis). Here, $c$ is the speed of light and $D$ is the
distance between the two LIGO sites. Because the measured
$\tau_\mathrm{M}$ has uncertainty $\delta\tau$, the ring has a finite
thickness. We assume that the probability distribution of the real time
delay, $\tau$, is a Gaussian around the measured time delay
$\tau_\mathrm{M}$ with the standard deviation $\delta\tau$. By changing
the variate from $\tau$ to $\theta$ using $\theta=\cos^{-1}\left(c
\tau/D\right)$, we get the SPDF for a LIGO event,
\begin{equation}
 S_\mathrm{GW}\left(\bi{r} ; \theta_\mathrm{ev},\delta\tau\right)=
  A_\mathrm{GW}\cdot\exp\left[-\frac{D^2\left(\cos\theta-\cos\theta_\mathrm{ev}\right)^2}{2\delta\tau^2
	c^2}\right],
\end{equation}
\begin{equation}
\theta=\cos^{-1}\left(\frac{\bi{r}\cdot\bi{l}}{\left|\bi{r}\right|\cdot\left|\bi{l}\right|}\right),
\end{equation}
where $\bi{l}$ is a vector parallel to the LIGO axis and $\theta$ is
the angle between $\bi{r}$ and the LIGO axis.
$S_\mathrm{GW}\left(\bi{r}; \theta_\mathrm{ev},\delta\tau\right)$ is normalized to unity over the whole sky by a normalization
factor $A_\mathrm{GW}$. An example of a LIGO event is shown in figure\,\ref{figure: SPDF}\,(a).

For the SPDF of an IceCube event we use a two-dimensional Gaussian
distribution on a sphere:

\begin{eqnarray}
S_\mathrm{\nu}\left(\bi{r}; \bi{r}_\mathrm{ev}, \sigma_\mathrm{\nu}\right)=A_\mathrm{\nu}\cdot\exp\left(\frac{-\psi^2}{2\sigma^2_\nu}\right), \\
\psi=\cos^{-1}\left(\frac{\bi{r}\cdot\bi{r}_\mathrm{ev}}{\left|\bi{r}\right|\cdot\left|\bi{r}_\mathrm{ev}\right|}\right),
\end{eqnarray}
where $\bi{r}_\mathrm{ev}$ is the vector representing the reconstructed
event direction and $\psi$ is the angle between $\bi{r}$ and
$\bi{r}_\mathrm{ev}$.  $A_\mathrm{\nu}$ is the normalization factor and $\sigma_\nu$
is the uncertainty of the reconstructed event direction. An example of
an IceCube event is shown in figure\,\ref{figure: SPDF}\,(b).

The distribution of background noise events is not uniform over the
sky. The background likelihood distribution (BLD) is a function of
 reconstructed event direction,
and it  returns a value proportional to the
likelihood of a background event coming from this direction.  The reconstructed event
direction is specified by a polar angle $\theta_\mathrm{ev}$ measured from the LIGO
axis for LIGO events and by a vector $\bi{r}_\mathrm{ev}$ for IceCube events.
There are two BLDs, $B_\mathrm{GW}(\theta_\mathrm{ev})$ and
$B_\mathrm{\nu}(\bi{r}_\mathrm{ev})$ corresponding to LIGO and IceCube detectors
respectively. BLDs are obtained from histograms of
reconstructed event directions, $\theta_\mathrm{ev}$ and
$\bi{r}_\mathrm{ev}$, for a large number of background events. The histograms
are converted to BLDs by normalizing them to 1 for integration over the whole sky. 

Finally,  the joint likelihood distribution of a combined
LIGO-IceCube event  is given by the following formula:
\begin{equation}
 L_\mathrm{comb}\left(\bi{r}\right)=\frac{S_\mathrm{GW}\left(\bi{r};\theta_\mathrm{ev},\delta\tau
						       \right)\cdot
  S_\mathrm{\nu}\left(\bi{r}; \bi{r}_\mathrm{ev}, \sigma_\mathrm{\nu}\right)}{B_\mathrm{GW}(\theta_\mathrm{ev})\cdot
  B_\mathrm{\nu}(\bi{r}_\mathrm{ev})}.
\end{equation}
 $L_\mathrm{comb}\left(\bi{r}\right)$ has a bright
 spot on the sky when the reconstructed directions of LIGO and IceCube
 events have good overlap.  We search for every direction on the sky
 and find  the direction $\bi{r}_\mathrm{max}$ which gives the maximum
 value
 $L_\mathrm{max}=L_\mathrm{comb}\left(\bi{r}_\mathrm{max}\right)=\mathrm{Max}\left[L_\mathrm{comb}\left(\bi{r}\right)\right]$.
 $L_\mathrm{max}$ is a good measure of spatial coincidence and
 $\bi{r}_\mathrm{max}$ is the most likely source direction.

In order to evaluate the statistical significance of a given
$L_\mathrm{max}$, we first calculate the background distribution
$P^\mathrm{BG}_{L_\mathrm{max}}(L_\mathrm{max})$ of $L_\mathrm{max}$
using a large number of background events.
$P^\mathrm{BG}_{L_\mathrm{max}}(L_\mathrm{max})$ gives the probability
of a time-coincident background event
 to have a particular $L_\mathrm{max}$.
Then the statistical significance of a 
combined event with $L_\mathrm{max}=L_\mathrm{ev}$ is
 estimated by the p-value  defined as follows:
\begin{equation}
 p=\int^\infty_{L_\mathrm{ev}}P^\mathrm{BG}_{L_\mathrm{max}}\left(L_\mathrm{max}\right)dL_\mathrm{max}.
\end{equation} 
The p-value gives the probability for a background
 combined event to have a value
$L_\mathrm{max}$ higher  than the $L_\mathrm{max}$ of the
 event ($L_\mathrm{ev}$)
being examined. Therefore, smaller p-values indicate
the candidate is less likely to be a background noise event. A detection
is declared if the p-value of a candidate is less than a certain
threshold value $p_0$, which is chosen according to the required
statistical significance for detections.

\section{Monte Carlo simulation}
\label{Sec: Monte Carlo Simulation} 

 The performance of our analysis
pipeline was demonstrated using Monte Carlo simulations. We first
generated a LIGO event list using 17.6 hours of LIGO-like data which has
similar statistical properties (such as standard deviation, glitch rate,
etc.)  to the real LIGO data during the fifth scientific run
(S5)\,\cite{2006CQGra..23..S03W}.  Using the statistics of LIGO events
obtained from this list (i.e. the event rate and the distributions of
$\tau$ and $\delta\tau$ used below), we generated a large number of
background LIGO events by Monte Carlo. For each event, a trigger time
was assigned randomly with the event rate of 13.4 events per
day, which is what we can reasonably expect from a real
detector on average. The arrival time difference $\tau$ between the
two LIGO sites was distributed uniformly between -10\,msec and 10\,msec.
The uncertainty $\delta\tau$ of the time difference was generated
following the gamma distribution below:
\begin{equation}
 P_{\delta\tau}(\delta\tau)=\frac{1}{b^a
 \Gamma\left(a\right)}\left(\delta\tau\right)^{a-1}e^{-\delta\tau/b},
\end{equation}
\[
 a=1.93,\quad b=4.41\times 10^{-4}.
\]
 This distribution was chosen by a fit to the histogram of $\delta\tau$
 obtained from the LIGO-like data.

Simulated IceCube events are distributed uniformly over the northern
hemisphere of the sky with an event rate of 2 events
per day. This event rate corresponds to the one obtained during the
operation of IceCube in its nine-string configuration
from June to November of 2006\,\cite{2007ICRC...IPC}.  No IceCube events
from the southern sky are generated because they are rejected by the
IceCube event reconstruction algorithm to avoid contamination by cosmic
ray muons. The uncertainty $\sigma_\nu$ of the event
direction is set to be a constant value of $2^\circ$, which is the
median angular reconstruction error of IceCube in the nine-string
configuration.

\begin{figure}[tbp]
\begin{center}
 
 \begin{minipage}[c]{6cm}
 \begin{center}
 \includegraphics[width=6cm]{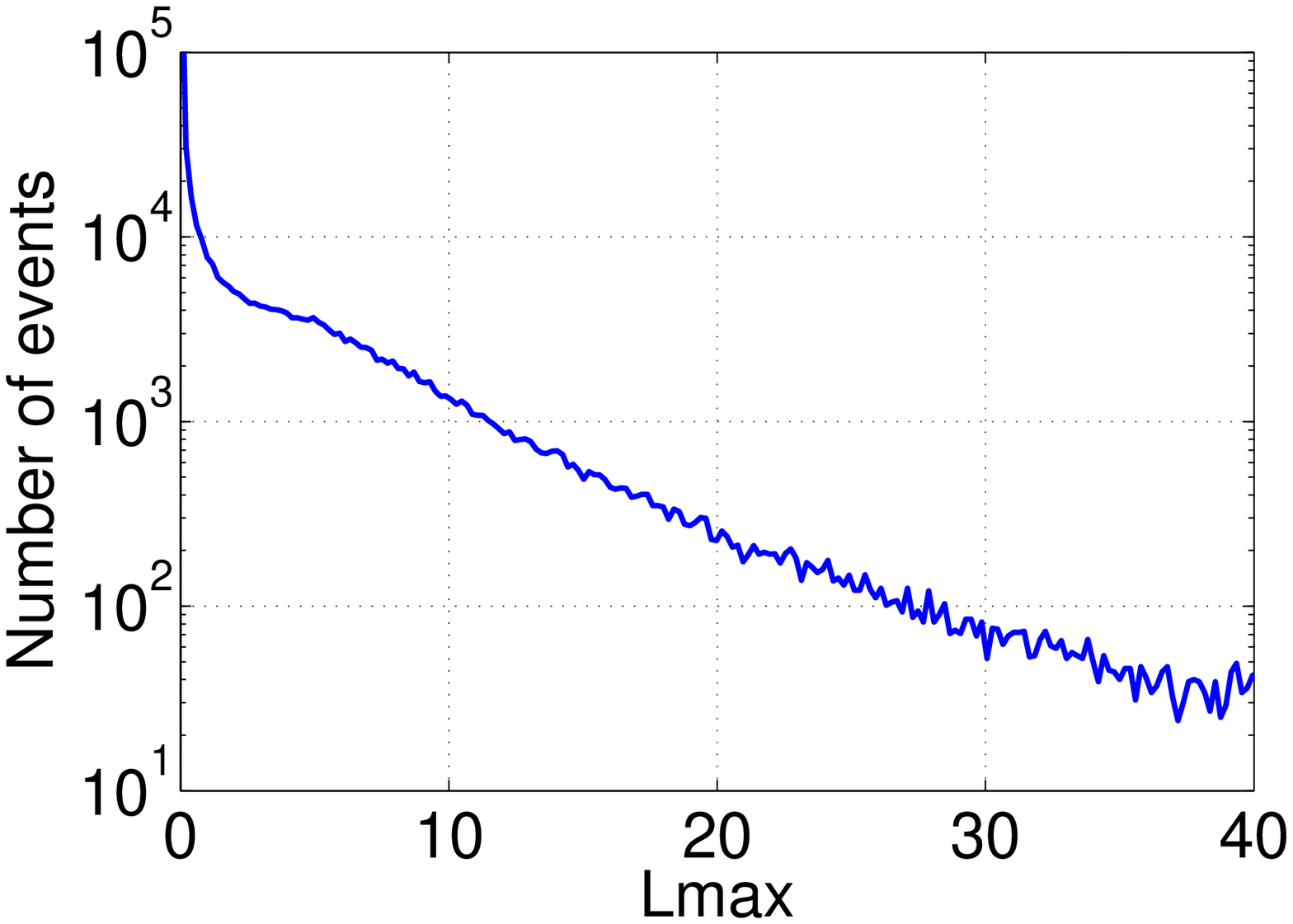}\\
(a)
  \end{center} 
 \end{minipage}
\begin{minipage}[c]{6cm}
 \begin{center}
 \includegraphics[width=6cm]{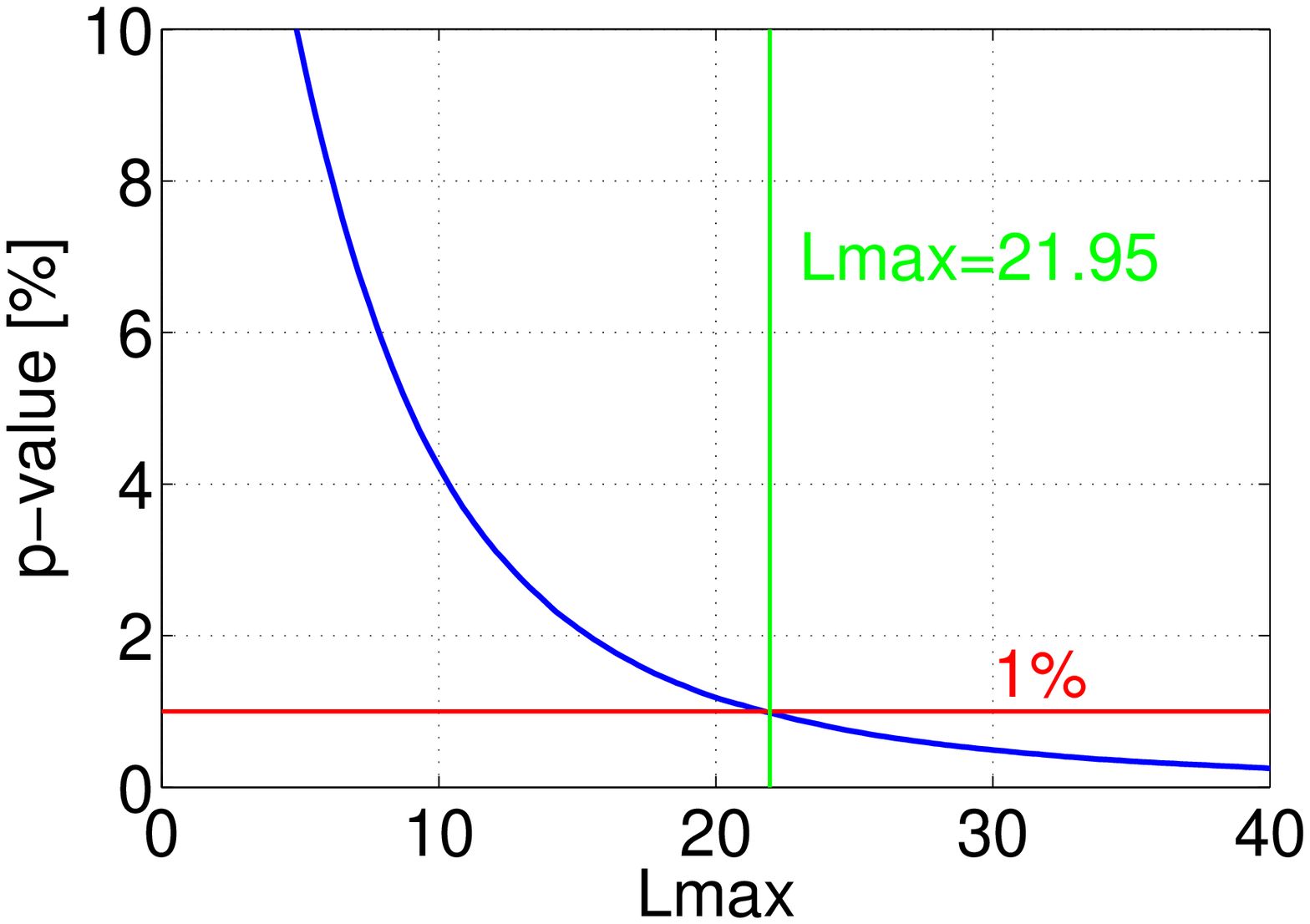}\\
(b)
 \end{center}
\end{minipage}
\caption{(a) The histogram of $L_\mathrm{max}$ for background
 events. (b) The plot of p-value as a function of $L_\mathrm{max}$. Less than 1\,\% of
background coincident events have an $L_\mathrm{max}$ value greater than 21.95. }  
\label{figure: Lmax and p-value}
\end{center}
\end{figure}

The simulated LIGO and IceCube events are fed into our analysis
pipeline. Figure\,\ref{figure: Lmax and p-value}\,(a) shows the
 distribution of
$L_\mathrm{max}$ for background coincident events. By integrating the
histogram, we get the relation between the p-value
for spatial coincidence and $L_\mathrm{max}$
(Figure\,\ref{figure: Lmax and p-value}\,(b)).  From this plot, we can
determine the detection threshold for $L_\mathrm{max}$. For example,
if our analysis requires that the p-value for spatial
coincidence be less than 1\,\%, then the $L_\mathrm{max}$ of the
combined event must be greater than the corresponding $L_\mathrm{max}$
value of 21.95.

\section{Discussion}
For each LIGO event, the expected number of IceCube
events found within a time window ($\pm T_\mathrm{W}$) is $2T_\mathrm{W}\cdot
R_\mathrm{\nu}$, where $R_\mathrm{\nu}$ is the event rate of
IceCube. Therefore, using the event rate $R_\mathrm{GW}$ of LIGO, the
 overall rate for
 LIGO-IceCube  time-coincident events can be calculated by
$2T_\mathrm{W}\cdot R_\mathrm{GW}\cdot R_\mathrm{\nu}$. 
 Using a p-value threshold $p_0$ for spatial coincidence, the FAR of this analysis method can be expressed
by the following formula,
\begin{equation}
\mathrm{FAR}= 2T_\mathrm{W}\cdot R_\mathrm{GW} \cdot R_\mathrm{\nu}\cdot p_0.
\end{equation}
More specifically in the case of the Monte Carlo simulation
explained in the previous section, the
FAR is given by the following formula,
\begin{equation}
\mathrm{FAR}=\frac{1}{435}\left(\frac{p_0}{1\,\%}\right)
 \left(\frac{T_\mathrm{W}}{1\,\mathrm{sec}}\right)\,\mathrm{[events /
 year]}.
\end{equation} 

The obtained FAR is 1 false alarm in 435 years for
one-second coincidence time window and spatial coincidence p-value
threshold of 1\,\%. If we allow a higher FAR, for example 1
event per 100 years used by SNEWS (SuperNova Early Warning
System)\,\cite{Antonioli04}, we can relax the LIGO or IceCube event selection
thresholds to search for weaker signals in the background
noise.

In the case of long GRBs, high energy neutrinos from relativistic shocks
are expected to be emitted  between a few
hours (for the internal shocks\,\cite{1997PhRvL..78.2292W, GUPTA07,
2006PhRvL..97e1101M}) to a few days (for the external
shocks\,\cite{2000..Wax..ApJ.Neut}) after gravitational wave emission
caused by core bounce.  In order to look for this type of
event, we have to use a large time window of
order of days.  In this case, the FAR may be unacceptably large because most LIGO
events will be able to find at least one companion IceCube event (and
vice versa) within a day. However, if the discrimination power of spatial
coincidence can be improved, this would offset the larger time overlap.
Such improvement would result from, for example, the continued
enlargement of the IceCube detector, or the addition of another
gravitational wave detector operating in conjunction with LIGO.
 On the other hand, the time coincidence is
effective to search for GW and neutrino bursts with small time delay.

Our method can also be applied to coincidence analyses with other
neutrino detectors such as Super-Kamiokande\,\cite{Fukuda03}, Lake
Baikal\,\cite{2006astro.ph..9743W}, Baksan\,\cite{1998PPN....29..254A}
etc, without  significant modification. Combinations
with the low energy-threshold detectors would enable us to search for
supernova events. Moreover, our method can be used with any
astronomical detectors which provide timing and source location
information of burst events. Coincidence search with a large number of
detectors will increase the confidence of detections.

We shall extend our method to include the VIRGO gravitational wave
detector. The use of three geographically separated interferometers will
enable us to constrain possible source locations of a gravitational wave
event to two points on the sky\,\cite{1989AsAph..217..381B}. Additionally, time
coincidence discrimination between VIRGO and LIGO interferometers will
further reduce the background event rate of the gravitational wave
detectors network.  Both of these changes in the time and spatial coincidence
rates will work together to provide a much lower FAR and/or better
sensitivity.

\ack 

The authors are grateful for the support of the United States National
Science Foundation under cooperative agreement PHY-04-57528 and Columbia
University in the City of New York. We are grateful to the LIGO
collaboration for their support. We are indebted to Jamie Rollins for
his useful comments on the manuscript.
The authors gratefully acknowledge the support of the United States
National Science Foundation for the construction and operation of the
LIGO Laboratory.
This work was also supported in part by the Office of Polar Programs of
the National Science Foundation and a Grant-in-Aid for Scientific
Research from the Ministry of Education, Science and Culture of Japan
through No.S 19104006.
This paper has been assigned LIGO Document Number LIGO-P070115-01-Z. 
\section*{References}
\bibliographystyle{myunsrt}
\bibliography{Manuscript}

\end{document}